\documentclass{ws-procs9x6}            
\usepackage{mathrsfs}
\usepackage{float}
\usepackage{hyperref}
\usepackage{url}
\usepackage{color}
\usepackage{xcolor}
\usepackage{epsfig}
\usepackage{epstopdf}
\DeclareGraphicsExtensions{.eps,.pdf,.png,.jpg,.jpeg}
\begin{document}
\title{Higgs dark energy in inert doublet model}

\author{Muhammad Usman$^*$ and Asghar Qadir$^\dagger$}

\address{School of Natural Sciences (SNS),
	National University of Sciences and Technology (NUST),
	Sector H-12, Islamabad 44000, Pakistan\\
	E-mail: muhammad\_usman\_sharif@yahoo.com$^*$, asgharqadir46@gmail.com$^\dagger$
}

\begin{abstract}
Scalar fields are among the possible candidates for dark energy. This paper is devoted to the scalar fields from the inert doublet model, where instead of one as in the standard model, two SU(2) Higgs doublets are used. The component fields of one SU(2) doublet ($\phi_1$) act in an identical way to the standard model Higgs while the component fields of the second SU(2) doublet ($\phi_2$) are taken to be the dark energy candidate (which is done by assuming that the phase transition in the field has not yet occurred
). It is found that one can arrange for late time acceleration (dark energy) by using an SU(2) Higgs doublet in the inert Higgs doublet model, whose vacuum expectation value is zero, in the quintessential regime.
\end{abstract}


\bodymatter
	\section{Introduction}
	Three components which contribute to the total energy density of the Universe are non-relativistic matter, relativistic matter and dark energy. The dark energy, which constitutes about $70\%$ of the present energy density of the Universe is thought to be providing the accelerated expansion to the Universe. Several possible candidates (explanations) of inflation and dark energy are discussed in the literature, a few of them are: the cosmological constant ($\Lambda$) \cite{Peebles,muhammadsami}; modified gravity \cite{Faraoni}\cite{Nojiri}; scalar field models (e.g. quintessence, K-essence, tachyon field, phantom (ghost) field \cite{trodden,0264-9381-19-17-311,PhysRevD.70.107301}, dilatonic dark energy, Chaplygin gas) \cite{muhammadsami} and vector fields 
	\cite{vectorinflation}.
	
	The homogeneous, isotropic Universe model is described by the Friedmann-Robertson-Walker (FRW) metric and its dynamics is described by the Friedmann equations which are
	\begin{eqnarray}
	H^2 &= \dfrac{1}{3}\rho-\dfrac{\kappa}{a^2}~,\qquad \label{1stFriedmannequation}
	\dfrac{\ddot{a}}{a} &= -\dfrac{1}{6}\left(1+3\omega_{\text{eff}}\right)\rho~, \label{2ndFriemannequation}
	\end{eqnarray}
	where $a$ is the scale factor, $H=\dot{a}/a$ is the Hubble parameter, $\rho$ is the total energy density, $\kappa$ is spatial curvature and $P$ is the pressure. In deriving the above equations, $\omega_{\text{eff}}=\Omega_{DE}\text{\space}\omega_{DE}+\Omega_{R}\text{\space}\omega_{R}+\Omega_{M}\text{\space}\omega_{M}$ and the barotropic equation of state $P=\omega\rho$ has been used. From eq. (\ref{2ndFriemannequation}), we see that $\ddot{a}>0$ when $\omega_{\text{eff}}<-\frac{1}{3}$.
	
	
	
	In this paper, we assume that the present state of the Universe is described by an inert Higgs doublet model. 
	If the phase transition in the second SU(2) doublet, $\phi_2$, (which has zero vacuum expectation value (vev)) has not occurred yet, then the component fields of the doublet can be considered as dark energy candidates. 
\section{The inert doublet model}
The electroweak symmetry in the standard model (SM) of Particle Physics is broken spontaneously by the non-zero vev of the Higgs field(s) via the Higgs mechanism. The Lagrangian which describes any model in Particle Physics is
\begin{equation}\begin{array}{rcl}\label{L}
\mathscr{L}=\mathscr{L}^{SM}_{gf}+\mathscr{L}_{Y}+\mathscr{L}_{Higgs}~.
\end{array}
\end{equation}
Here, $\mathscr{L}^{SM}_{gf}$ is the $SU_{C}(3){\otimes}SU_{L}(2){\otimes}U_{Y}(1)$ is the SM interaction of the fermions and gauge bosons (force carriers)
, $\mathscr{L}_{Y}$ is the Yukawa interaction of fermions with the Higgs doublet which has non zero vev in inert doublet model
and $\mathscr{L}_{Higgs}$ is the Higgs field Lagrangian where
\begin{align}\begin{split}\label{LH}
\mathscr{L}_{Higgs}&=({D_1}_{\mu}\phi_1)^\dagger ({D_1}^{\mu}\phi_1) + ({D_2}_{\mu}\phi_2)^\dagger ({D_2}^{\mu}\phi_2) \\ &+\left [ {\chi}({D_1}_{\mu}{\phi}_{1})^{\dagger}({D_2}^{\mu}{\phi}_{2})+{\chi^{*}}({D_2}_{\mu}{\phi}_{2})^{\dagger}({D_1}^{\mu}{\phi}_{1}) \right ]-V(H),
\end{split}\end{align}
where the first three terms describe the kinetic energy and $V_{H}$ is the potential of the Higgs field(s)
\begin{align}
V_H &=\text{\space} V_1+V_2+V_{int} ~, \label{V12int}\\
\begin{split}\label{VH}
&=\text{\space}\rho_1\exp(\Lambda_1 m_{11}^2\phi_1^\dagger\phi_1)+\rho_2\exp(\Lambda_2 m_{22}^2\phi_2^\dagger\phi_2)+\rho_3\exp{\Big(}\dfrac{1}{2}\Lambda_3\lambda_1(\phi_1^\dagger\phi_1)^2{\Big)} \\ &\text{\space\space\space} +\rho_4\exp{\Big(}\dfrac{1}{2}\Lambda_4\lambda_2(\phi_2^\dagger\phi_2)^2{\Big)}+{m_{12}^2}(\phi_1^\dagger \phi_2)+{{m_{12}^2}^*}(\phi_2^\dagger \phi_1)
+\lambda_3 (\phi_1^\dagger\phi_1)(\phi_2^\dagger\phi_2)
\\ &\text{\space\space\space}
+\lambda_4(\phi_1^\dagger\phi_2)(\phi_2^\dagger\phi_1)+\dfrac{1}{2}{\big[}\lambda_5(\phi_1^\dagger\phi_2)^2 +\lambda_5^* (\phi_2^\dagger\phi_1)^2{\big]}
+\lambda_6(\phi_1^\dagger\phi_1)(\phi_1^\dagger\phi_2)
\\ &\text{\space\space\space}
+\lambda_6^*(\phi_1^\dagger\phi_1)(\phi_2^\dagger\phi_1)+\lambda_7(\phi_2^\dagger\phi_2)(\phi_1^\dagger\phi_2)+\lambda_7^*(\phi_2^\dagger\phi_2)(\phi_2^\dagger\phi_1)~,
\end{split}
\end{align}
where $V_1$ and $V_2$ in eq. (\ref{V12int}) are the Lagrangian of the Higgs field $\phi_1$ (given by the first two terms of the RHS of eq. (\ref{VH})) and $\phi_2$ (given by the $3^{rd}$ and the $4^{th}$ terms of the RHS of eq. (\ref{VH})) respectively, $V_{int}$ is the interaction Lagrangian of the fields $\phi_1$ and $\phi_2$ given by the remaining terms. The covariant derivative and the Higgs doublets are defined as
$${D_i}_\mu=\partial_\mu+\dot{\iota}\dfrac{g_i}{2}\sigma_j {W^j}_\mu+\dot{\iota}\dfrac{g_i'}{2}B_\mu ~,$$
$$
\phi_{i}=
\begin{bmatrix}
\phi^{+}_{i} \\
\eta_i + \dot{\iota}\chi_i +\nu_i \\
\end{bmatrix},
\text{\qquad \quad \qquad}
\phi_{i}^{\dagger}=
\begin{bmatrix}
\phi^{-}_{i} & \eta_i - \dot{\iota}\chi_i +\nu_i
\end{bmatrix}.
$$
Where $g_i$ and $g_i^{'}$ are the SU(2) and U(1) couplings of the Higgs doublets respectively, $\sigma_j$ are the Pauli matrices, ${W^j}_\mu$ and $B_\mu$ are the generators of SU(2) and U(1) group. The dimensions of different parameters are
$$[\rho_i]^{-1}=[\Lambda_i]=[L]^4,\text{\space} [m_{ii}^2]=[L]^{-2},\text{\space} [\phi_i]=[L]^{-1}\text{\space and \space}[\lambda_i]=[L]^0.$$
Here ``$L$'' denotes the length. The fields $\phi^{+}_{i}$, $\phi^{-}_{i}$, $\eta_i$ and $\chi_i$ are the hermitian Higgs fields ($\phi^{\pm}_{i}$ are charged whereas other fields are neutral), $\nu_i$ is the vev of the doublet $\phi_i$.
The vev of the potential is found by taking
\begin{equation}\label{extremaconditions}
\dfrac{\partial V_H}{\partial \phi_1} {\bigg|}_{\substack{
		\phi_1=\left\langle \phi_1\right\rangle \\
		\phi_2=\left\langle \phi_2\right\rangle
	}}=\dfrac{\partial V_H}{\partial \phi_1^\dagger}\bigg |_{\substack{
	\phi_1=\left\langle \phi_1\right\rangle \\
	\phi_2=\left\langle \phi_2\right\rangle
}}=0 
\text{\space\space and \space} 
\dfrac{\partial V_H}{\partial \phi_2}\bigg |_{\substack{
\phi_1=\left\langle \phi_1\right\rangle \\
\phi_2=\left\langle \phi_2\right\rangle
}}=\dfrac{\partial V_H}{\partial \phi_2^\dagger}\bigg |_{\substack{
\phi_1=\left\langle \phi_1\right\rangle \\
\phi_2=\left\langle \phi_2\right\rangle
}}=0~.
\end{equation}
One solution of the above equations, by truncating the potential eq. (\ref{VH}) upto forth order in the $Z_2$ symmetry conservative case, for $\nu_1$ and $\nu_2$ is
\begin{eqnarray}
\nu_1^2 &= -\dfrac{\rho_1 \Lambda_1 m_{11}^2}{\lambda_1^{'}} ~, \text{\quad\quad\quad\space\space\space} \nu_2^2 &= 0 ~, \label{inertvacuum2}
\end{eqnarray}
where 
$\lambda_1^{'}=\dfrac{1}{2}\left(\rho_3\Lambda_3\lambda_1+\rho_1(\Lambda_1 m_{11}^2)^2\right)$ and $\lambda_2^{'}=\dfrac{1}{2}\left(\rho_4\Lambda_4\lambda_2+\rho_2(\Lambda_2 m_{22}^2)^2\right)$.
The masses of the Higgs fields in this vacuum are
\begin{equation}\begin{array}{rcl}\label{inertvacuumHiggsmasses}
&& m_{\eta_1}^2=2\lambda_1^{'}\nu_1^2~,\qquad m_{\eta_2\space,\space\chi_2}^2=\rho_2\Lambda_2m_{22}^2+\dfrac{\lambda_3+\lambda_4\pm\text{Re}(\lambda_5)}{2}\nu_1^2~,
\\ &&
m_{\phi_2^{\pm}}^2=\rho_2\Lambda_2m_{22}^2+\dfrac{\lambda_3}{2}\nu_1^2~,
\end{array}
\end{equation}
where $\nu_1=\dfrac{1}{\sqrt[4]{2{G_F}^2}}\approx246$GeV. The Higgs vacuum energy of the potential given by eq. (\ref{VH}) is 
\begin{equation}\label{Evacinertmodel}
\begin{array}{rcl}
E^{'}_{vac}&=&\rho_2+\rho_4+\rho_1 \exp\left(\dfrac{1}{2}m_{11}^2\Lambda_1\nu_1^2\right)+\rho_3 \exp\left(\dfrac{1}{8}\lambda_1\Lambda_3\nu_1^4\right).
\end{array}
\end{equation}
If we require that the phase transitions in the doublet $\phi_1$ (which is acting as the SM Higgs doublet) occur at the same temperature as in the SM Higgs doublet we obtain
\begin{equation}\label{phasetransitionsconstraint}
2\lambda_3+\lambda_4=0.7869~,
\end{equation}
\section{Higgs field(s) as dark energy field(s)}
\begin{figure}[H]
	\centering
	\includegraphics[scale=.2]{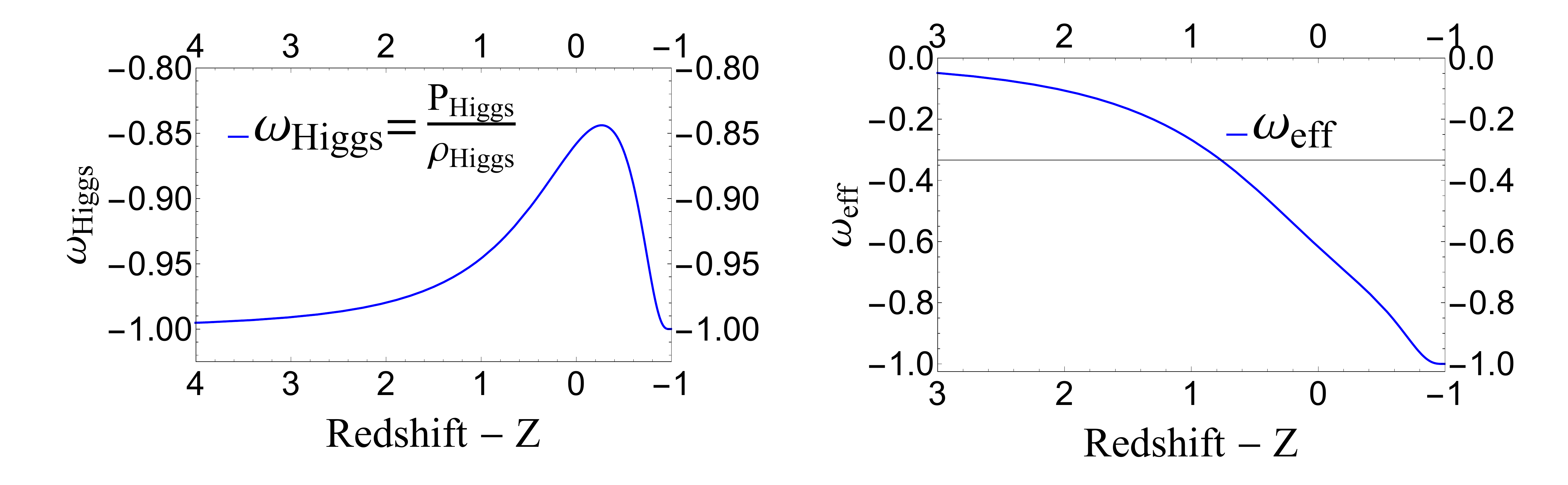}
	\caption{Equation of state $\omega_{Higgs}$ and $\omega_{eff}$ as a function of reshift}
	\label{fig:OmegaHiggs}
\end{figure}

\begin{figure}[H]
	\centering
	\includegraphics[scale=0.2]{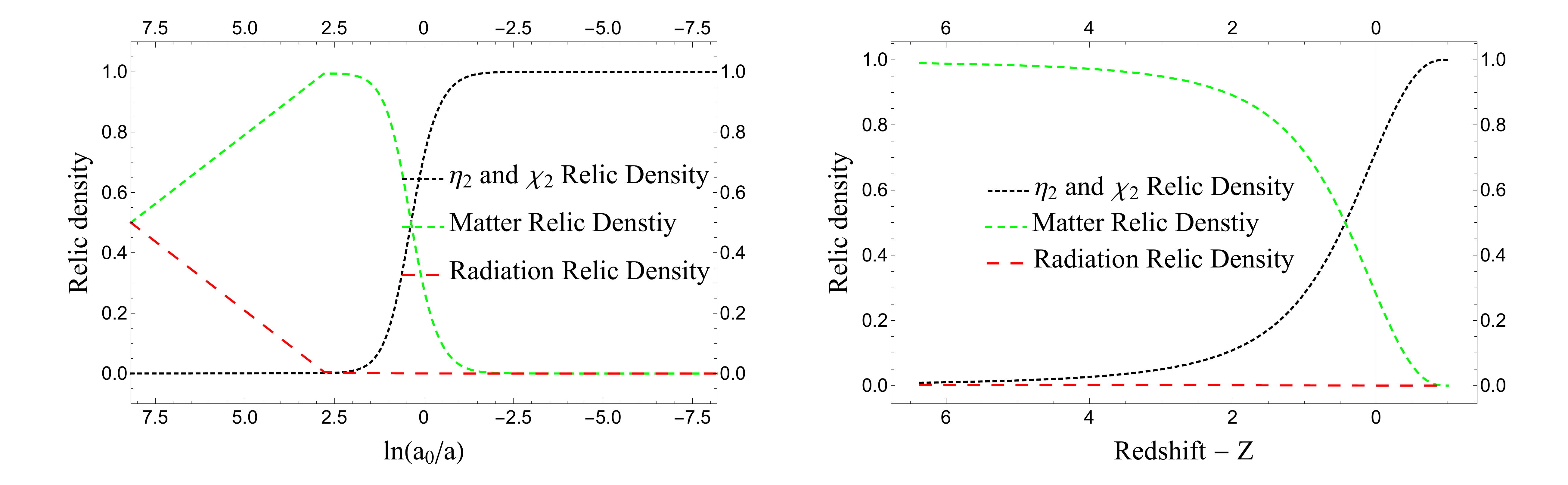}
	\caption{Relic densities as a function of $ln$[$a_0/a$].}
	\label{fig:relicdensities1}
\end{figure}

By the second Friedmann eq. (\ref{2ndFriemannequation}) the accelerated expansion will occur when $\omega_{\text{eff}}<-\frac{1}{3}$.
For the cosmological evolution of the fields $\eta_2$, $\chi_2$ and $\phi_2^c$, the Euler-Lagrange equations of motion of the fields are solved with the Friedmann equations numerically in the flat Universe ($\kappa=0$). The initial conditions used are ${\eta_2}_{ini}=M_P$, ${\chi_2}_{ini}=M_P$, ${\phi_2^c}_{ini}=0$, ${\dot{\eta_2}}_{ini}=0$, ${\dot{\chi_2}}_{ini}=0$ and ${\dot{\phi_2^c}}_{ini}=0$.
The masses of the Higgs bosons in the analysis are taken to be
\begin{equation*}
m_{\eta_2,\chi_2}=6.925\times 10^{-61}\text{GeV,\quad} m_{\phi_2^\pm}=154.305\text{GeV},
\end{equation*}
so as to get the evolution of the energy densities as observed. Since the observed vacuum energy is so small $\mathcal{O}(10^{-121}E_P/L_P^3)$, it needs for very small values of parameters to achieve over-damped oscillations (required to ensure that $\omega_{\text{eff}}$ become less than $-1/3$ only once on the cosmological scale). 
Thus, there is also a fine tuning problem related to the smallness of the field mass in our model.
The numerical solution obtained is depicted graphically in Figs. \ref{fig:OmegaHiggs} and \ref{fig:relicdensities1},
During the evolution, equation of state parameter $\omega_{Higgs}$ starts to increasing from $-1$ as shown in Fig. \ref{fig:OmegaHiggs}. In the very late (future) Universe ($Z\leq0$), a period with $\omega=-1$ is reachieved to give an exponentially accelerating Universe.

The $\omega_{\text{eff}}$ in Fig. \ref{fig:OmegaHiggs} starts from $\approx0.167$ (set by initial conditions $\Omega_{{Higgs}_{int}}=0$ and $\Omega_{{NR}_{int}}=\Omega_{R_{int}}=0.5$; where NR stands for non-relativistic matter and R for relativistic matter) and decreases as the energy density of relativistic matter decreases. This and before is the time period when non-relativistic matter dominates and expanding Universe decelerates as matter domination pulls everything inwards more than the outward Higgs negative pressure. After that $\omega_{\text{eff}}$ starts to decrease as the non-relativistic energy density decreases and Higgs relic energy density increases as shown in Fig. \ref{fig:relicdensities1}, this time and afterwards Higgs negative pressure starts to dominates for forever and $\omega_{\text{eff}}$ eventually settles down to $-1$.

Figure \ref{fig:relicdensities1} is plotted for the evolution of the relic density of the non-relativistic matter, relativistic matter and the dark energy Higgs fields against $\text{ln}(a_0/a(t))$. The figure shows that only now the dark energy Higgs has started to dominate in the critical density while in the past non-relativistic matter was dominating. The dark energy Higgs becomes dominant at $Z=6$ and will dominate afterwards. At $Z\approx 0.3235$ the non-relativistic matter and relativistic matter energy densities are equal. For $Z > 0.3235$ non-relativistic matter is dominating and for $Z < 0.3235$ dark energy Higgs started to dominate causing the accelerated expansion of the Universe.
\section{Conclusions}\label{Conclusion}

In this article we assumed that dark energy actually is some form of scalar field(s) which is (are) present in the so called inert Higgs doublet model.
The masses $m_i$ were chosen arbitrary to get $\Omega_{Higgs}\approx0.7$, to avoid oscillations in the fields we have to take $\lambda_3+\lambda_4\pm\lambda_5\approx\mathcal{O}(10^{-125})$ and thus the masses of second Higgs doublet's fields very small to ensure that we get $\omega_{\text{eff}}=-1/3$ only once in the history of the Universe.


One thing that remains important to check in all extensions of the SM is whether the Higgs potential contains the vacuum instability or not? 
The answer to the question for our model is that although it contains the vacuum instability, due to the coupling of the second Higgs with the SM Higgs $\mathcal{O}(10^{-126})$, the RGEs running of the SM Higgs will not be effected. We expect the vacuum instability to occur at approximately the same scale as it occurs in the SM.


When describing a model for accelerated expansion of the Universe, it becomes essential to compare it with the $\Lambda$CDM model. 
In comparing our model with the $\Lambda$CDM, we noted that on the cosmological scale our model is quite different from the standard $\Lambda$CDM in low redshift era. In our model $\omega_{Higgs}$ is not constant over redshift, as shown in Fig. \ref{fig:OmegaHiggs}, giving the approximate value of $\omega_{Higgs}$ to be $-0.858$ at $Z=0$, whereas in $\Lambda$CDM one expects to get $\omega_{\text{DE}}$ to be $-1$ at all redshifts. Thus, our model can be distinguished from the $\Lambda$CDM via the variation of $\omega_{Higgs}$  from $-1$ on the cosmological scale. 
In concluding, since we get $\omega_{eff}<-1/3$ after solving the Euler-Lagrange equations numerically, the proposed Higgs field could cause the current observed accelerated expansion.

\section*{Acknowledgment}

This work is supported by {\textit{National University of Sciences and Technology (NUST), Sector H-12 Islamabad 44000, Pakistan}} and \emph{Higher education commission (HEC) of Pakistan} under the project no. NRPU-3053.

\nocite{*}
\bibliographystyle{ws-procs9x6} 
\bibliography{Muhammad-Usman}

\providecommand{\noopsort}[1]{}\providecommand{\singleletter}[1]{#1}%
\begin{thebibliography}{1}

\bibitem{Peebles}
P.~J.~E. Peebles and B.~Ratra,  (2002).

\bibitem{muhammadsami}
E.~J. Copeland, M.~Sami and S.~Tsujikawa,  (2006).

\bibitem{Faraoni}
T.~P. Sotiriou and V.~Faraoni, {\em Rev.\ Mod.\ Phys.} {\bf 82}, p. 451
  (2010).

\bibitem{Nojiri}
S.~Nojiri and S.~D. Odintsov, {\em Phys.\ Rep.\ 505} {\bf 59}  (2011).

\bibitem{trodden}
S.~M. Carroll, M.~Hoffman and M.~Trodden, {\em Phys.\ Rev.\ D} {\bf 68}, p.
  023509  (2003).

\bibitem{0264-9381-19-17-311}
V.~K. Onemli and R.~P. Woodard, Super-acceleration from massless, minimally
  coupled $\phi^4$, {\em Classical and Quantum Gravity} {\bf 19}, p. 4607
  (2002).

\bibitem{PhysRevD.70.107301}
V.~K. Onemli and R.~P. Woodard, Quantum effects can render $\omega<-1$ on
  cosmological scales, {\em Phys. Rev. D} {\bf 70}, p. 107301 (Nov 2004).

\bibitem{vectorinflation}
A.~Golovnev, V.~Mukhanov and V.~Vanchurin, {\em JCAP06} {\bf 009}  (2008).

\end{thebibliography}
\end{document}